\documentclass[conference]{IEEEtran}
\usepackage{graphicx} 
\usepackage{multirow}
\usepackage{colortbl}
 \usepackage{booktabs}
 \usepackage[table]{xcolor}
 \usepackage{threeparttable}
 \usepackage{balance}
 \usepackage{url} 
\definecolor{good}{RGB}{198,239,206}
\definecolor{moderate}{RGB}{255,235,156}
\definecolor{bad}{RGB}{255,199,206}

\usepackage[table]{xcolor}

\definecolor{lightgreen}{RGB}{220,255,220}
\definecolor{lightyellow}{RGB}{255,245,200}
\definecolor{lightorange}{RGB}{255,220,180}
\definecolor{lightred}{RGB}{255,200,200}
\usepackage{booktabs}
\usepackage{xcolor}
\usepackage{threeparttable}

\usepackage[table]{xcolor}
\usepackage{booktabs}
\usepackage{caption}

\definecolor{moderate}{RGB}{255,245,200}

 \definecolor{good}{RGB}{198,239,206}
 \definecolor{bad}{RGB}{255,199,206}
\usepackage[letterpaper, left=0.625in, right=0.625in, bottom=1in, top=0.75in]{geometry}
\usepackage[T1]{fontenc}

\usepackage{xcolor}
\usepackage{fancyhdr}

\fancypagestyle{firstpagefooter}{
  \fancyhf{}
  \fancyfoot[C]{\textcolor{red}{\footnotesize
  © 2026 IEEE. Accepted for publication in the Proceedings of IEEE LCN 2026. The final version will appear in IEEE Xplore.}}

}

\title{Energy-Aware System-Level Evaluation of Post-Quantum TLS on Embedded User Equipment over a Disaggregated 5G Network}
\author{Sanzida Hoque, Abdullah Aydeger\\
Florida Institute of Technology, Melbourne, FL, USA, 32901\\
\protect Emails: shoque2023@my.fit.edu, aaydeger@fit.edu}

\date{April 2026}

\begin{document}

\maketitle
\thispagestyle{firstpagefooter} 
\begin{abstract}
The transition to quantum-resistant security is a critical priority for the next generation of mobile networks, particularly within the disaggregated architecture of 5G. This paper presents an energy-aware system-level evaluation of Post-Quantum Cryptography (PQC) integrated into the Transport Layer Security (TLS) handshake on embedded User Equipment (UE). Using Raspberry Pi 5s as representative embedded processing platforms, we evaluate the performance of NIST-standardized combinations of classical and post-quantum signature and key exchange mechanisms (KEM), incorporating direct on-device power measurements to estimate per-handshake energy consumption. Results 
\textcolor{black}{experimentally validate} a strong coupling between latency and energy consumption, indicating that execution time is the dominant contributor to energy cost. Hash-based signature schemes incur up to 4x higher latency and 2x energy compared to lattice-based alternatives, while the impact of KEMs is comparatively smaller. The analysis further reveals that overall system performance is primarily constrained by cryptographic computation and concurrency-induced contention rather than network transport effects. These findings provide practical guidance for PQC deployment in mobile environments and demonstrate that lattice-based signatures offer a more favorable balance between security, efficiency, and scalability for 5G systems.
\end{abstract}

\section{Introduction}
The emergence of quantum computing poses a fundamental threat to widely deployed public-key cryptographic systems, including Rivest–Shamir–Adleman (RSA) and elliptic curve cryptography (ECC). These schemes form the foundation of secure communication protocols such as Transport Layer Security (TLS), which are extensively used in modern cellular networks \cite{lastre2025evaluating}. In 5G systems, secure communication between user equipment (UE), the radio access network, and the core infrastructure is essential for ensuring data confidentiality and integrity. Consequently, the transition to post-quantum cryptography (PQC) has become a critical requirement for future-proofing mobile communication systems \cite{hoque2024exploring}.

Recent standardization efforts by the National Institute of Standards and Technology (NIST) \cite{nist_pqc_overview, nist_ir_8545_2025} have identified several promising PQC algorithms, including lattice-based and hash-based schemes. While these algorithms provide resistance against quantum adversaries, they introduce additional computational complexity compared to classical approaches, this raises concerns regarding their deployment in resource-constrained environments such as mobile devices, where latency, energy efficiency, and processing capability are tightly constrained.

Existing research on PQC performance has largely focused on isolated cryptographic benchmarks or protocol-level evaluations in controlled environments. While valuable, these studies often overlook system-level interactions in real-world deployments, including network stacks, protocol implementations, and device-level power consumption. In 5G, where strict performance and energy requirements must be met, these factors are crucial to assessing the practical feasibility of PQC adoption.

To address this gap, this paper presents a system-level evaluation of PQC-enabled TLS handshakes on an end-to-end 5G testbed integrating embedded UE devices, an emulated NG-RAN, and a standards-compliant 5G core. This setup enables end-to-end experimentation under controlled yet representative conditions. Furthermore, the use of onboard power monitoring allows direct observation of UE-side energy consumption during cryptographic operations, providing insights that are not captured by conventional benchmarking approaches.
The main contributions of this work are as follows:
   \begin{itemize}
    \item We conduct a system-level evaluation of classical and post-quantum TLS handshakes on an end-to-end 5G testbed, using embedded UE integrated with disaggregated radio access and core network components.
    
    \item We employ on-device Power Management Integrated Circuit (PMIC) measurements to approximate UE-side power and energy consumption, enabling practical and reproducible evaluation without external instrumentation.
    
    \item We analyze the relationship between latency and energy consumption, showing that execution time is a primary contributor to energy cost in PQC-enabled TLS.
    
    \item We examine the impact of cryptographic primitives on system performance, observing that signature algorithms contribute the majority of latency and energy overhead, with hash-based schemes incurring higher cost than lattice-based alternatives.
    
    \item We provide a concurrency-driven evaluation of post-quantum TLS, quantifying how increasing numbers of client UEs impact energy consumption, system stress, and stability on edge-class hardware, identifying distinct performance trends and the onset of saturation at high load. 
    
    \item We assess the relative contributions of computation and networking to end-to-end performance, showing that cryptographic processing dominates under the evaluated conditions.
\end{itemize}
The remainder of this paper is organized as follows. Section \ref{sec:rw} reviews related work. Section \ref{sec:SD} describes the architecture and design of the system. Section \ref{sec:results} presents experimental methodology, the results, and discussion. Finally, Section \ref{sec:conclusion} concludes the paper and outlines future research directions.

\section{Related Work}
\label{sec:rw}
The transition toward PQC has been accelerated by recent standardization activities. In 2024, NIST finalized the first set of post-quantum cryptographic standards, \textcolor{black}{including the Module-Lattice-Based Key-Encapsulation Mechanism (ML-KEM, FIPS 203 \cite{fips203_mlkem}), Module-Lattice-Based Digital Signature Algorithm (ML-DSA, FIPS 204 \cite{fips204_mldsa}), and Stateless Hash-Based Digital Signature Algorithm (SLH-DSA, FIPS 205 \cite{fips205_slh_dsa}) for quantum-resistant key establishment and digital signatures.} This standardization has motivated extensive research on integrating PQC into transport protocols, particularly TLS~1.3, where cryptographic primitives directly impact handshake latency, message size, and deployment compatibility. 

\textit{PQC Integration in TLS Protocols.} A growing body of work evaluates PQC within transport-layer protocols. Recent studies show that ML-KEM-based TLS introduces moderate overhead, primarily due to increased key sizes and message payloads rather than computational complexity \cite{montenegro2025comparative}. Other work demonstrates that hardware-aware optimizations can significantly improve PQC-TLS handshake performance \cite{zheng2024faster}. Prior work has evaluated PQC-enabled TLS in constrained IoT settings, including embedded-device energy consumption \cite{tasopoulos2023energy} and TLS tunneling \cite{barton2022post}. Kampanakis et al. \cite{kampanakis2024impact} further show that PQC handshake overhead has limited end-to-end impact in stable, high-bandwidth networks as application data increases. \textcolor{black}{However, these studies remain largely protocol-level or controlled-environment evaluations and do not assess PQC-TLS in realistic 5G deployments, where protocol overhead, network behavior, device performance, and energy consumption interact.}

\textit{PQC Tooling and Industry Deployment.} To support such experimentation, the Open Quantum Safe ecosystem has emerged as a widely adopted platform for PQC prototyping. The liboqs library enables integration of PQC algorithms into TLS stacks such as OpenSSL, wolfSSL, and BoringSSL. Industry deployments further demonstrate the practical relevance of PQC-enabled TLS. For example, Google has introduced hybrid post-quantum KEM in Chrome \cite{chromium_pqc}, while Cloudflare reports that a significant portion of Internet traffic is already protected using post-quantum key agreement \cite{cloudflare2025pqc}. These efforts highlight the rapid transition toward PQC in real-world systems, particularly for mitigating “harvest-now, decrypt-later” threats. Similarly, industry analyses, including those by Akamai and others, highlight practical challenges in integrating hybrid PQC into TLS, such as increased handshake sizes, compatibility constraints, and client-server interoperability issues \cite{schaumann2025pqc_tls}. While these efforts provide valuable insights into deployment feasibility at Internet scale, they primarily focus on compatibility and protocol behavior, and do not address device-level energy characteristics or system-level evaluation in mobile network environments.

\textit{PQC on Embedded and Edge Platforms.} Another line of work investigates PQC performance on embedded and edge platforms. Patterson et al. \cite{patterson2025energy} propose an energy measurement framework for PQC key generation on Raspberry Pi devices, demonstrating that PQC schemes can be competitive with classical cryptography under certain conditions. However, their analysis is limited to isolated cryptographic operations and does not consider end-to-end communication. Similarly, recent studies show that PQC-enabled TLS can be deployed on embedded devices with manageable overhead \cite{ince2026hybrid}. Recent benchmarking efforts further evaluate PQC-enabled TLS in embedded and networked environments, focusing on certificate size and handshake latency \cite{sim2025pqc}. Despite these contributions, such evaluations are typically conducted in standalone or conventional network environments and do not capture device-level energy behavior or system-level effects. Liu et al. \cite{liu2025post} demonstrate PQC migration on a physical 5G testbed, highlighting challenges such as larger ciphertexts, buffer constraints, and latency overheads in protocols like TLS and SUCI. In contrast, our work focuses on system stress and energy behavior under concurrent client UEs. 

\textit{PQC in 5G and System-Level Studies.} More recent work has begun to bridge the gap between protocol-level and system-level evaluation. 
\textcolor{black}{Our prior work \cite{hoque2025analysis} demonstrated the feasibility of PQC-enabled TLS in a 5G UE-to-UE communication setting using UERANSIM and Open5GS, focusing on system-level metrics such as latency, CPU utilization, and bandwidth overhead. This paper extends that work by shifting the focus to an energy-aware evaluation on resource-constrained Raspberry Pi 5 devices, jointly analyzing power consumption, energy cost, thermal behavior, and concurrency effects.}
Other studies explore integrating PQC into 5G systems \cite{hoque2026post, hoque2024post}, including modifications to core network signaling, demonstrating that PQC can be incorporated into mobile networks with acceptable overhead \cite{faval2026empowering}. Scalise et al. explored integrating PQC key encapsulation mechanisms into 5G core networks, demonstrating minimal impact on latency and bandwidth during VNF-to-VNF communication \cite{scalise2024applied}.  However, these works primarily focus on network and protocol performance rather than device-level energy and thermal behavior.

In summary, prior work has independently addressed (i) PQC-enabled TLS performance, (ii) embedded PQC execution, and (iii) \textcolor{black}{5G testbed-based performance evaluation}, but these dimensions remain disconnected: protocol-level studies overlook mobile network effects, embedded evaluations omit end-to-end communication, and 5G studies do not characterize device-level power and thermal behavior. This paper bridges these gaps by evaluating PQC-enabled TLS on physical embedded UE over a disaggregated 5G data path, jointly characterizing latency, energy, instantaneous power, and thermal response across the full set of NIST-standardized PQC KEM and signature schemes.

\section{System Architecture and Design}
\label{sec:SD}
\begin{figure}[htbp]
    \centering
    \includegraphics[width=1\linewidth]{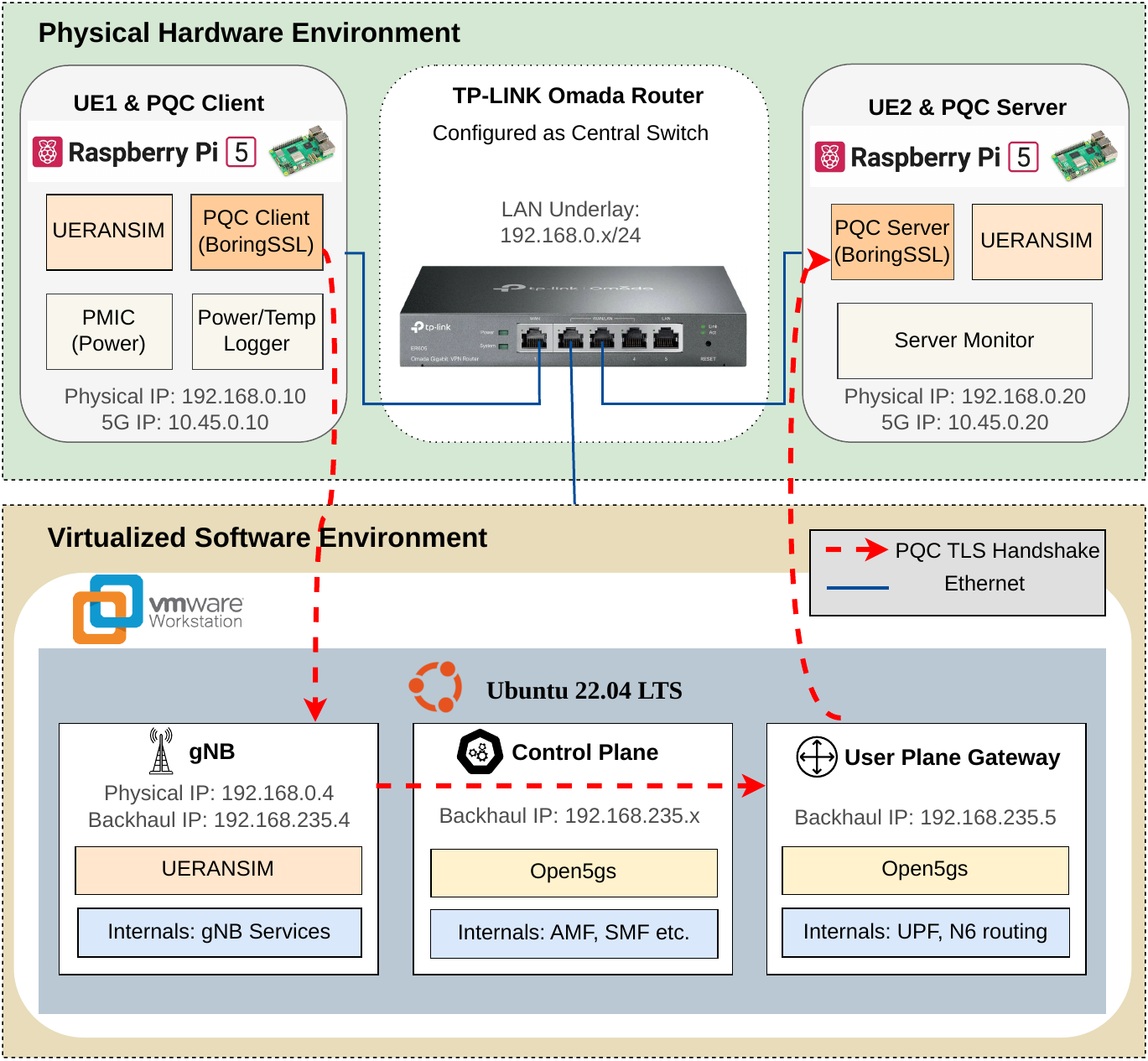}
    \caption{Disaggregated 5G PQC Testbed Architecture. The red-dashed path denotes the logical end-to-end PQC TLS handshake overlay, while the blue solid lines indicate the hardware-level Ethernet connections.}
    \label{fig:sa}
\end{figure}

This work presents a disaggregated PQC-enabled 5G testbed designed to precisely measure the computational, thermal, and energy characteristics of PQC algorithms as illustrated in Fig \ref{fig:sa}. 
The system is composed of two tightly integrated domains: (i) a physical hardware layer, where PQC operations are executed on real devices, and (ii) a virtualized 5G network layer, which emulates the Radio Access Network (RAN) and Core Network functions. This separation enables independent control of computation and networking variables by executing PQC operations on dedicated physical hardware while confining network functions to a virtualized environment, thereby eliminating resource contention and scheduler-induced interference between the two domains.  
\subsection{Physical Hardware Layer} 
The hardware layer comprises two embedded nodes, both implemented on Raspberry Pi 5 devices and each operating as a 5G UE: (i) UE1 acts as the TLS client, initiating PQC-enabled handshakes; (ii) UE2 acts as the TLS server, terminating the handshakes and processing application data. Each node is instrumented to monitor system behavior during execution. Power consumption is obtained from onboard power interfaces, and temperature is collected from integrated sensors. These measurements enable correlation between cryptographic operations and device-level energy and thermal responses.
The nodes are interconnected via an Ethernet underlay network that provides connectivity to the virtualized environment.


\subsection{Virtualized Software Layer}
The virtualized software layer emulates a disaggregated 5G network. The network functions are deployed in a disaggregated manner, separating control-plane signaling from user-plane data forwarding in accordance with contemporary 5G architectural principles. The virtualized RAN component (gNB) provides the access interface, bridging the physical nodes to the core network. The core network itself is logically divided into: (i) a control plane, responsible for signaling, authentication, and session management, and
(ii) a user plane, responsible for forwarding application data. This separation ensures that control signaling does not interfere with user-plane measurements, thereby preserving the integrity of PQC performance evaluation.

\subsection{Data Flow and Operation} The end-to-end communication follows a structured multi-stage process. The UE client node generates PQC-enabled TLS traffic, which is forwarded to the virtualized access network. The traffic is then encapsulated within a tunneling protocol and transported through the virtualized core network. Upon reaching the user-plane gateway, the packet is decapsulated and forwarded to the destination node in the physical layer.

This design ensures that the network path reflects 5G data-path behavior, including GTP-U encapsulation and UPF forwarding, while the cryptographic operations remain fully observable on physical hardware. The radio access is emulated by UERANSIM rather than transmitted over-the-air; this study targets the cryptographic-cost and data-path dimensions, which are independent of PHY/MAC effects.

\section{Experimental Evaluation}
\label{sec:results}
\subsection{Experimental Setup}
Experiments are conducted on the testbed described in Section \ref{sec:SD}. Both embedded nodes are implemented on Raspberry Pi 5 with \textcolor{black}{8GB} RAM devices and operate as 5G UEs using UERANSIM. At the application layer, one node acts as a client and the other as a server, executing PQC-enabled TLS sessions using a BoringSSL integrated with the liboqs to support PQC.

The virtualized 5G network is deployed using UERANSIM for the access network (gNB) and Open5GS for core network functionality, hosted on a single-core Intel64 system operating at 1.7 GHz via a VMware-based platform.
Each experiment consists of repeated TLS session establishments under identical conditions. TP-Link Omada ER605 router serves as the central switching fabric, providing the physical underlay network. 

The network follows a multi-tier addressing scheme: the LAN Underlay (192.168.0.x) provides the physical interconnect via the Omada switch, the RAN Backhaul (192.168.235.x) facilitates gNB-to-Core communication, and the 5G User Plane (10.45.0.x) encapsulates the PQC handshake. All reported TLS handshakes are exchanged over the UE-assigned 5G user-plane addresses (10.45.0.x), traversing the UERANSIM gNB and Open5GS UPF path, rather than directly over the physical LAN underlay, and they are averaged across multiple runs to ensure consistency and reliability. Although the host running the virtualized 5G plane is modestly provisioned, the consistently low and stable RTT observed across all concurrency levels (Section IV-C5) confirms that the virtualized network plane is not a measurement bottleneck.
\begin{table*}[t]
\centering
\caption{End-to-End Performance Across Concurrency Levels (Client-Observed Latency, Server-Observed Throughput, and Energy)}
\label{tab:e2e}
\scriptsize
\begin{threeparttable}
\begin{tabular}{lcccccccccccccccccl}
\toprule
& \multicolumn{5}{c}{Latency (ms)} 
& \multicolumn{5}{c}{Throughput (KB/s)} 
& \multicolumn{5}{c}{Energy (mJ/conn)} 
& Category \\
\cmidrule(lr){2-6} \cmidrule(lr){7-11} \cmidrule(lr){12-16}
Configuration 
& C1 & C4 & C10 & C20 & C40 
& C1 & C4 & C10 & C20 & C40 
& C1 & C4 & C10 & C20 & C40 
& Class \\
\midrule

P-256 + SLH-DSA & 202 & 506 & 1277 & 2355 & 4733 & 225 & 376 & 399 & 447 & 446 & 213 & 238 & 200 & 200 & 200 & \cellcolor{bad}High Cost \\
X25519 + SLH-DSA & 199 & 470 & 1202 & 2404 & 4800 & 226 & 420 & 432 & 446 & 445 & 215 & 219 & 208 & 219 & 199 & \cellcolor{bad}High Cost \\
ML-KEM + SLH-DSA & 194 & 459 & 1201 & 2362 & 4750 & 233 & 448 & 452 & 462 & 462 & 233 & 225 & 200 & 217 & 203 & \cellcolor{bad}High Cost \\
HQC + SLH-DSA & 203 & 481 & 1262 & 2524 & 4500 & 264 & 495 & 476 & 495 & 504 & 250 & 200 & 207 & 198 & 336 & \cellcolor{bad}Worst Case \\

\midrule

P-256 + Falcon & 107 & 118 & 306 & 632 & 1282 & 39 & 148 & 160 & 157 & 168 & 188 & 139 & 151 & 142 & 144 & \cellcolor{good}Efficient \\
X25519 + Falcon & 111 & 133 & 370 & 756 & 1550 & 39 & 135 & 135 & 143 & 143 & 188 & 148 & 139 & 147 & 139 & \cellcolor{moderate}Moderate \\
ML-KEM + Falcon & 104 & 120 & 312 & 636 & 1274 & 52 & 348 & 212 & 218 & 223 & 143 & 126 & 142 & 141 & 140 & \cellcolor{good}Best Hybrid \\
HQC + Falcon & 139 & 208 & 546 & 1105 & 2260 & 88 & 242 & 263 & 296 & 288 & 197 & 155 & 145 & 157 & 142 & \cellcolor{moderate}Heavy \\

\midrule

ML-KEM + ML-DSA & 108 & 118 & 315 & 648 & 1325 & 94 & 347 & 366 & 374 & 389 & 180 & 126 & 145 & 145 & 140 & \cellcolor{good}Balanced \\
HQC + ML-DSA & 137 & 205 & 552 & 1104 & 2180 & 119 & 351 & 353 & 377 & 416 & 194 & 160 & 150 & 146 & 148 & \cellcolor{moderate}Heavy \\
P-256 + ML-DSA & 109 & 122 & 322 & 671 & 1345 & 81 & 291 & 308 & 316 & 316 & 223 & 151 & 140 & 139 & 140 & \cellcolor{good}Efficient \\
X25519 + ML-DSA & 115 & 142 & 385 & 795 & 1600 & 80 & 260 & 266 & 268 & 272 & 131 & 139 & 139 & 143 & 146 & \cellcolor{moderate}Moderate \\

\midrule

P-256 + P-256 & 106 & 120 & 306 & 626 & 1260 & 18 & 68 & 75 & 78 & 88 & 176 & 133 & 141 & 145 & 139 & \cellcolor{good}Baseline \\
X25519 + P-256 & 110 & 132 & 363 & 743 & 1527 & 18 & 63 & 66 & 75 & 79 & 177 & 149 & 141 & 132 & 139 & \cellcolor{moderate}Moderate \\
ML-KEM + P-256 & 106 & 120 & 305 & 629 & 1265 & 32 & 116 & 130 & 138 & 151 & 151 & 146 & 140 & 139 & 141 & \cellcolor{good}Efficient Hybrid \\
HQC + P-256 & 135 & 207 & 549 & 1111 & 2250 & 76 & 194 & 205 & 238 & 262 & 194 & 144 & 149 & 144 & 152 & \cellcolor{moderate}Heavy \\

\bottomrule
\end{tabular}

\begin{tablenotes}
\footnotesize
\item ML-KEM denotes ML-KEM-512; ML-DSA denotes ML-DSA-44; Falcon (aka FN-DSA) denotes Falcon512; SLH-DSA denotes SPHINCS+ SHA2-128f-simple.
\item Concurrency levels (C1-C40) indicate the number of simultaneous client UEs.
\end{tablenotes}

\end{threeparttable}
\end{table*}

\begin{table*}[t]
\centering
\caption{System Resource Utilization Across Concurrency Levels (CPU Utilization, Server Power Consumption, and Temperature)}
\label{tab:resource}
\scriptsize
\begin{threeparttable}
\begin{tabular}{lcccccccccccccccc}
\toprule
& \multicolumn{5}{c}{CPU (\%)} 
& \multicolumn{5}{c}{Server Avg Power (mW)} 
& \multicolumn{5}{c}{Server Peak Temp (°C)} \\
\cmidrule(lr){2-6} \cmidrule(lr){7-11} \cmidrule(lr){12-16}
Configuration 
& C1 & C4 & C10 & C20 & C40 
& C1 & C4 & C10 & C20 & C40 
& C1 & C4 & C10 & C20 & C40 \\
\midrule

P-256 + SLH-DSA 
& \cellcolor{moderate}6 & \cellcolor{moderate}6 & \cellcolor{bad}21 & \cellcolor{bad}31 & \cellcolor{bad}36
& \cellcolor{good}1245 & \cellcolor{good}1308 & \cellcolor{moderate}1846 & \cellcolor{bad}2292 & \cellcolor{bad}2464
& 62 & 66 & 67 & 69 & 69 \\

X25519 + SLH-DSA 
& \cellcolor{moderate}5 & \cellcolor{bad}18 & \cellcolor{bad}24 & \cellcolor{bad}34 & \cellcolor{bad}35
& \cellcolor{good}1257 & \cellcolor{moderate}1855 & \cellcolor{moderate}1991 & \cellcolor{bad}2484 & \cellcolor{moderate}2194
& 62 & 65 & 68 & 69 & 70 \\

ML-KEM + SLH-DSA 
& \cellcolor{moderate}8 & \cellcolor{moderate}7 & \cellcolor{bad}22 & \cellcolor{bad}22 & \cellcolor{bad}33
& \cellcolor{good}1375 & \cellcolor{good}1306 & \cellcolor{moderate}1912 & \cellcolor{moderate}1972 & \cellcolor{bad}2344
& 63 & 65 & 67 & 68 & 69 \\

HQC + SLH-DSA 
& \cellcolor{bad}11 & \cellcolor{bad}22 & \cellcolor{bad}22 & \cellcolor{bad}31 & \cellcolor{bad}35
& \cellcolor{moderate}1510 & \cellcolor{moderate}1828 & \cellcolor{moderate}1845 & \cellcolor{moderate}2193 & \cellcolor{bad}2424
& 62 & 65 & 68 & 69 & 69 \\

\midrule

P-256 + Falcon 
& \cellcolor{good}0.3 & \cellcolor{good}0.9 & \cellcolor{good}1.1 & \cellcolor{good}1.5 & \cellcolor{good}1.5
& \cellcolor{good}1284 & \cellcolor{moderate}1809 & \cellcolor{moderate}1984 & \cellcolor{bad}2393 & \cellcolor{bad}2376
& 61 & 64 & 67 & 68 & 68 \\

X25519 + Falcon 
& \cellcolor{good}0.7 & \cellcolor{good}1.6 & \cellcolor{moderate}2.3 & \cellcolor{moderate}2.3 & \cellcolor{moderate}3.1
& \cellcolor{good}1320 & \cellcolor{moderate}1657 & \cellcolor{moderate}1958 & \cellcolor{moderate}1972 & \cellcolor{bad}2274
& 60 & 64 & 68 & 68 & 69 \\

ML-KEM + Falcon 
& \cellcolor{good}0.4 & \cellcolor{good}1.1 & \cellcolor{good}1.1 & \cellcolor{good}1.1 & \cellcolor{good}1.3
& \cellcolor{good}1366 & \cellcolor{moderate}1896 & \cellcolor{bad}2206 & \cellcolor{bad}2249 & \cellcolor{bad}2430
& 62 & 66 & 67 & 68 & 68 \\

HQC + Falcon 
& \cellcolor{good}1.5 & \cellcolor{moderate}3.2 & \cellcolor{moderate}5.3 & \cellcolor{moderate}5.3 & \cellcolor{moderate}7.0
& \cellcolor{good}1343 & \cellcolor{moderate}1692 & \cellcolor{moderate}2097 & \cellcolor{moderate}2177 & \cellcolor{bad}2482
& 60 & 65 & 68 & 69 & 70 \\

\midrule

ML-KEM + ML-DSA 
& \cellcolor{good}0.2 & \cellcolor{good}1.1 & \cellcolor{good}1.5 & \cellcolor{good}1.3 & \cellcolor{good}1.5
& \cellcolor{good}1223 & \cellcolor{moderate}1897 & \cellcolor{bad}2297 & \cellcolor{moderate}2156 & \cellcolor{bad}2331
& 60 & 66 & 68 & 68 & 68 \\

HQC + ML-DSA 
& \cellcolor{good}1.1 & \cellcolor{moderate}3.8 & \cellcolor{moderate}6.5 & \cellcolor{moderate}7.2 & \cellcolor{moderate}7.6
& \cellcolor{good}1207 & \cellcolor{moderate}1793 & \cellcolor{bad}2335 & \cellcolor{bad}2465 & \cellcolor{bad}2545
& 60 & 66 & 68 & 69 & 69 \\

P-256 + ML-DSA 
& \cellcolor{good}0.4 & \cellcolor{good}1.3 & \cellcolor{good}1.3 & \cellcolor{good}1.5 & \cellcolor{good}1.0
& \cellcolor{good}1338 & \cellcolor{moderate}1962 & \cellcolor{moderate}1984 & \cellcolor{moderate}2094 & \cellcolor{moderate}1714
& 60 & 64 & 67 & 68 & 68 \\

X25519 + ML-DSA 
& \cellcolor{good}0.9 & \cellcolor{moderate}2.3 & \cellcolor{moderate}2.4 & \cellcolor{moderate}3.7 & \cellcolor{moderate}3.7
& \cellcolor{good}1241 & \cellcolor{moderate}1830 & \cellcolor{moderate}1887 & \cellcolor{bad}2446 & \cellcolor{bad}2450
& 61 & 65 & 67 & 68 & 68 \\

\midrule

P-256 + P-256 
& \cellcolor{good}0.2 & \cellcolor{good}0.5 & \cellcolor{good}0.7 & \cellcolor{good}0.8 & \cellcolor{good}1.2
& \cellcolor{good}1336 & \cellcolor{moderate}1703 & \cellcolor{moderate}1879 & \cellcolor{moderate}1967 & \cellcolor{bad}2440
& 61 & 65 & 67 & 69 & 68 \\

X25519 + P-256 
& \cellcolor{good}0.7 & \cellcolor{moderate}1.9 & \cellcolor{moderate}2.6 & \cellcolor{moderate}2.8 & \cellcolor{moderate}3.1
& \cellcolor{good}1329 & \cellcolor{moderate}1851 & \cellcolor{moderate}2162 & \cellcolor{bad}2262 & \cellcolor{bad}2442
& 60 & 65 & 67 & 68 & 69 \\

ML-KEM + P-256 
& \cellcolor{good}0.2 & \cellcolor{good}0.7 & \cellcolor{good}0.9 & \cellcolor{good}0.9 & \cellcolor{good}1.0
& \cellcolor{good}1242 & \cellcolor{moderate}1959 & \cellcolor{bad}2228 & \cellcolor{moderate}2192 & \cellcolor{bad}2349
& 61 & 65 & 67 & 69 & 68 \\

HQC + P-256 
& \cellcolor{good}1.4 & \cellcolor{moderate}4.9 & \cellcolor{moderate}5.4 & \cellcolor{moderate}6.8 & \cellcolor{moderate}6.0
& \cellcolor{good}1332 & \cellcolor{moderate}1969 & \cellcolor{moderate}2172 & \cellcolor{bad}2517 & \cellcolor{bad}2295
& 60 & 66 & 67 & 69 & 69 \\

\bottomrule
\end{tabular}

\begin{tablenotes}
\footnotesize
\item CPU: green $<$ 2\%, yellow 2-10\%, red $>$10\%.
\item Power: green $<$1500 mW, yellow 1500-2200 mW, red $>$2200 mW.
\item Temperature not color-coded (all within safe operating range).
\end{tablenotes}

\end{threeparttable}
\end{table*}

\begin{table*}[t]
\centering
\caption{Network Stability and Client-Side Impact (Retransmission, Round Trip Time, Client UE Power Draw)}
\label{tab:network}
\scriptsize
\begin{threeparttable}
\begin{tabular}{lcccccccccccccccc}
\toprule
& \multicolumn{5}{c}{Retransmissions} 
& \multicolumn{5}{c}{RTT (ms)} 
& \multicolumn{5}{c}{UE Peak Power (mW)} \\
\cmidrule(lr){2-6} \cmidrule(lr){7-11} \cmidrule(lr){12-16}
Configuration 
& C1 & C4 & C10 & C20 & C40 
& C1 & C4 & C10 & C20 & C40 
& C1 & C4 & C10 & C20 & C40 \\
\midrule

P-256 + SLH-DSA 
& \cellcolor{good}0 & \cellcolor{moderate}4 & \cellcolor{moderate}2 & \cellcolor{good}0 & \cellcolor{moderate}1
& 4.7 & 5.3 & 5.3 & 4.9 & 4.9
& \cellcolor{good}2775 & \cellcolor{moderate}5172 & \cellcolor{moderate}5328 & \cellcolor{bad}6717 & \cellcolor{bad}8026 \\

X25519 + SLH-DSA 
& \cellcolor{moderate}1 & \cellcolor{moderate}1 & \cellcolor{good}0 & \cellcolor{good}0 & \cellcolor{moderate}2
& 4.7 & 4.8 & 4.8 & 4.8 & 4.9
& \cellcolor{moderate}3774 & \cellcolor{moderate}5524 & \cellcolor{bad}6200 & \cellcolor{moderate}5599 & \cellcolor{bad}7534 \\

ML-KEM + SLH-DSA 
& \cellcolor{good}0 & \cellcolor{good}0 & \cellcolor{moderate}2 & \cellcolor{moderate}1 & \cellcolor{moderate}1
& 4.6 & 4.8 & 4.9 & 4.9 & 4.9
& \cellcolor{good}3195 & \cellcolor{good}3188 & \cellcolor{moderate}5329 & \cellcolor{bad}7785 & \cellcolor{bad}7851 \\

HQC + SLH-DSA 
& \cellcolor{good}0 & \cellcolor{moderate}2 & \cellcolor{bad}12 & \cellcolor{bad}26 & \cellcolor{bad}7
& 5.4 & 5.5 & 5.6 & 5.6 & 5.7
& \cellcolor{good}3305 & \cellcolor{moderate}3625 & \cellcolor{bad}6041 & \cellcolor{bad}6066 & \cellcolor{bad}6429 \\

\midrule

P-256 + Falcon 
& \cellcolor{good}0 & \cellcolor{good}0 & \cellcolor{moderate}4 & \cellcolor{moderate}4 & \cellcolor{bad}20
& 13.3 & 13.9 & 15.9 & 15.9 & 15.7
& \cellcolor{moderate}4329 & \cellcolor{good}3406 & \cellcolor{bad}6529 & \cellcolor{good}3473 & \cellcolor{bad}7753 \\

X25519 + Falcon 
& \cellcolor{good}0 & \cellcolor{moderate}4 & \cellcolor{bad}14 & \cellcolor{bad}26 & \cellcolor{bad}54
& 12.8 & 13.8 & 16.0 & 16.0 & 15.9
& \cellcolor{good}2962 & \cellcolor{moderate}4633 & \cellcolor{good}3432 & \cellcolor{moderate}5175 & \cellcolor{moderate}5672 \\

ML-KEM + Falcon 
& \cellcolor{good}0 & \cellcolor{moderate}1 & \cellcolor{moderate}3 & \cellcolor{bad}6 & \cellcolor{bad}15
& 11.4 & 8.9 & 14.6 & 14.4 & 14.1
& \cellcolor{moderate}4943 & \cellcolor{good}3186 & \cellcolor{good}3336 & \cellcolor{moderate}5653 & \cellcolor{bad}6853 \\

HQC + Falcon 
& \cellcolor{good}0 & \cellcolor{moderate}1 & \cellcolor{moderate}4 & \cellcolor{bad}12 & \cellcolor{bad}22
& 10.8 & 13.1 & 13.1 & 13.2 & 13.2
& \cellcolor{good}3403 & \cellcolor{moderate}5452 & \cellcolor{bad}8258 & \cellcolor{moderate}4349 & \cellcolor{moderate}4379 \\

\midrule

ML-KEM + ML-DSA 
& \cellcolor{good}0 & \cellcolor{moderate}1 & \cellcolor{moderate}4 & \cellcolor{bad}6 & \cellcolor{bad}17
& 8.6 & 8.9 & 10.4 & 10.7 & 10.7
& \cellcolor{good}2681 & \cellcolor{good}3185 & \cellcolor{moderate}4190 & \cellcolor{good}3357 & \cellcolor{moderate}5552 \\

HQC + ML-DSA 
& \cellcolor{moderate}3 & \cellcolor{good}0 & \cellcolor{moderate}4 & \cellcolor{bad}9 & \cellcolor{bad}28
& 8.6 & 10.6 & 10.8 & 10.6 & 10.6
& \cellcolor{good}3403 & \cellcolor{moderate}3678 & \cellcolor{moderate}4237 & \cellcolor{moderate}4201 & \cellcolor{bad}6276 \\

P-256 + ML-DSA 
& \cellcolor{good}0 & \cellcolor{moderate}2 & \cellcolor{moderate}1 & \cellcolor{bad}13 & \cellcolor{bad}30
& 9.4 & 10.0 & 11.7 & 11.7 & 11.6
& \cellcolor{good}2726 & \cellcolor{good}3233 & \cellcolor{moderate}5233 & \cellcolor{good}3444 & \cellcolor{moderate}4754 \\

X25519 + ML-DSA 
& \cellcolor{good}0 & \cellcolor{moderate}4 & \cellcolor{bad}7 & \cellcolor{bad}24 & \cellcolor{bad}51
& 9.0 & 9.8 & 11.7 & 11.6 & 11.6
& \cellcolor{good}2726 & \cellcolor{good}3117 & \cellcolor{moderate}3510 & \cellcolor{good}3481 & \cellcolor{moderate}3618 \\

\midrule

P-256 + P-256 
& \cellcolor{good}0 & \cellcolor{moderate}1 & \cellcolor{moderate}2 & \cellcolor{moderate}2 & \cellcolor{bad}12
& 15.6 & 16.7 & 18.3 & 18.3 & 17.9
& \cellcolor{good}2884 & \cellcolor{moderate}3536 & \cellcolor{good}3310 & \cellcolor{moderate}5335 & \cellcolor{good}3557 \\

X25519 + P-256 
& \cellcolor{good}0 & \cellcolor{moderate}1 & \cellcolor{moderate}4 & \cellcolor{moderate}4 & \cellcolor{bad}20
& 14.9 & 16.5 & 18.7 & 18.7 & 18.4
& \cellcolor{moderate}5434 & \cellcolor{good}3186 & \cellcolor{moderate}5613 & \cellcolor{good}3534 & \cellcolor{moderate}5200 \\

ML-KEM + P-256 
& \cellcolor{good}0 & \cellcolor{good}0 & \cellcolor{moderate}4 & \cellcolor{bad}11 & \cellcolor{bad}22
& 13.2 & 14.2 & 16.1 & 16.0 & 15.7
& \cellcolor{good}2617 & \cellcolor{good}3154 & \cellcolor{moderate}5264 & \cellcolor{moderate}4947 & \cellcolor{good}3618 \\

HQC + P-256 
& \cellcolor{good}0 & \cellcolor{moderate}1 & \cellcolor{bad}6 & \cellcolor{bad}26 & \cellcolor{bad}35
& 12.9 & 15.5 & 15.5 & 15.6 & 15.6
& \cellcolor{moderate}4804 & \cellcolor{moderate}3811 & \cellcolor{moderate}4254 & \cellcolor{moderate}4213 & \cellcolor{bad}6926 \\

\bottomrule
\end{tabular}

\begin{tablenotes}
\footnotesize
\item Retransmissions: green = 0, yellow = 1-5, red = $>$5.
\item UE peak power: green $<$3500 mW, yellow 3500-6000 mW, red $>$6000 mW.
\item RTT not color-coded due to negligible variation.
\end{tablenotes}

\end{threeparttable}
\end{table*}
\subsection{Metrics and Methodology}
\label{subsec:metrics_methodology}

We evaluate TLS handshake performance across heterogeneous cryptographic suites in a 5G edge deployment. Our methodology captures end-to-end performance, system resource utilization, and network/client-side effects under controlled multi-UE concurrency. 

\subsubsection{\textbf{Workload and Concurrency Model}}

We model realistic access scenarios by representing concurrency as multiple client UEs simultaneously initiating TLS handshakes toward a single edge server. This experiment isolates cryptographic processing load from RAN-level multi-UE scheduling effects, allowing the study to focus on the computational and energy cost of PQC under aggregated load.

Concurrency $C$ denotes the number of independent \textcolor{black}{BoringSSL} client UEs issuing handshake requests in parallel. We evaluate $C \in \{1,4,10,20,40\}$, chosen to reflect distinct operating regimes:

\begin{itemize}
    \item \textit{$C=1$ (Baseline):} Single UE, capturing pure cryptographic cost without contention.
    \item \textit{$C=4$ (Optimal Parallelism):} Matches the 4-core architecture of the server CPU, representing ideal hardware utilization.
    \item \textit{$C=10$ (Contention Onset):} Exceeds core count, introducing scheduling overhead and queueing.
    \item \textit{$C=20$ (Saturation):} System operates near maximum processing capacity.
    \item \textit{$C=40$ (Stress Limit):} Extreme load, exposing stability limits and failure behavior.
\end{itemize}

Each UE performs 50 TLS handshake attempts per run. Under high concurrency, some handshake attempts may not complete successfully due to system overload. We therefore distinguish between (i) attempted handshakes, defined as the total requests issued by all UEs, and (ii) completed handshakes, defined as the successfully processed connections at the server. All performance metrics (latency, throughput, and energy per connection) are computed using only completed handshakes, ensuring consistent comparison across configurations.

\textcolor{black}{Throughout this paper, power and energy are used with distinct meanings: power refers to instantaneous system load, whereas energy represents consumption over a given duration and reflects the cost of completing a TLS handshake.}





\subsubsection{\textbf{End-to-End Performance Metrics}}
These metrics capture the user-perceived performance of TLS handshakes under concurrent UE load.

Latency (ms):
Measured at the client UE as the \textcolor{black}{elapsed} time between handshake initiation and successful completion. Application-layer timestamps are used to ensure end-to-end coverage. \textcolor{black}{Same time-stamping procedure is used for all configurations to ensure fair comparison.} Reported values are averaged across completed handshakes. 

Throughput (KB/s):
Measured at the server as the total volume of handshake-related data processed per second. This includes key exchange messages, certificates, and signature payloads, which vary significantly across cryptographic schemes.

Energy per Connection (mJ/conn):
Derived from server-side power measurements. Average power over the experiment duration is used to estimate total energy consumption, which is then normalized by the number of completed handshakes.

\subsubsection{\textbf{Server-Side Resource Metrics}}
These metrics quantify computational load, energy consumption, and thermal behavior of the edge server.

CPU Utilization (\%):
Measured as average CPU usage across all cores during the experiment. This reflects the computational demand imposed by concurrent cryptographic operations and system overhead.

Server Average Power (mW):
Computed from PMIC telemetry by averaging power readings over the experiment duration. This captures steady-state energy consumption under load.

Server Peak Temperature ($^\circ$C):
Maximum observed CPU temperature during execution. This metric is used to verify whether the system approaches thermal throttling limits.

\subsubsection{\textbf{Network and Client-Side Metrics}}

These metrics capture network stability and client-side computational impact, enabling separation of communication and processing bottlenecks.

Retransmissions:
Measured using TCP statistics at the server. This metric indicates packet loss or delays caused by congestion or processing bottlenecks.

Round-Trip Time (RTT, ms):
Measured at the transport layer. Across all experiments, RTT remains nearly constant, indicating stable network conditions and confirming that performance degradation is not network-induced.

Client Peak Power (mW):
Maximum instantaneous power observed at the UE during handshake execution. This captures burst computational demand associated with cryptographic operations.
\subsubsection{\textbf{Measurement Limitations and Validity}}
While the proposed measurement framework provides comprehensive visibility into system behavior, several limitations should be noted.

Power measurements are obtained via the onboard PMIC, which reports estimated voltage and current values. As a result, absolute power may incur an error margin of approximately $\pm$5-10\%, and short-lived transient spikes may not be fully captured due to the sampling interval (100~ms). Similarly, client-side peak power relies on device-level telemetry, which may smooth high-frequency variations.

CPU utilization and latency measurements are subject to operating system scheduling and timer resolution \cite{etsion2003effects}, introducing minor variability (typically within 1-2~ms), particularly at low concurrency levels.

Retransmission statistics are derived from transport-layer counters and do not explicitly separate network losses from processing-induced delays. However, the consistently low and stable RTT across all experiments indicates that retransmissions are primarily driven by computational contention rather than network impairments.

Finally, experiments are conducted in a controlled testbed with stable networking conditions and active cooling. While this ensures reproducibility, it does not capture all environmental variability present in real deployments.

Despite these limitations, the results remain robust. All experiments are repeated \textcolor{black}{twice} under identical conditions with consistent trends across runs. \textcolor{black}{Our objective is comparative evaluation across cryptographic algorithms executed on identical hardware under identical conditions. Therefore, relative differences are more important than absolute values.} The observed differences across cryptographic configurations, often exceeding 2$\times$, significantly outweigh measurement uncertainty. Additionally, thermal measurements confirm operation below throttling thresholds, ensuring that performance degradation reflects system-level and computational effects rather than hardware constraints.
\subsection{Results Analysis}
\label{subsec:results}

This section presents a detailed analysis of system behavior under increasing concurrency, focusing on latency, energy consumption, computational load, network behavior, and hardware characteristics. Results are derived from Tables~\ref{tab:e2e}, \ref{tab:resource}, and \ref{tab:network}. Across these results, three trends dominate. First, signature choice has a stronger effect on latency, energy, CPU utilization, and retransmissions than KEM choice. Second, ML-KEM paired with Falcon or ML-DSA remains close to classical performance across the evaluated concurrency range. Third, SLH-DSA pushes the system toward saturation at high concurrency, increasing latency and CPU utilization while reducing completed handshakes in the worst case.

\subsubsection{Latency Scaling with Concurrency}

Latency increases with concurrency across all evaluated configurations, with distinct scaling behavior depending on the cryptographic primitives. SLH-DSA-based configurations exhibit the steepest increase, rising from approximately 195-205~ms at $C=1$ to 4700-4800~ms at $C=40$, corresponding to more than a $20\times$ increase.

In contrast, lattice-based configurations demonstrate significantly more moderate scaling. For example, ML-KEM + Falcon increases from approximately 104~ms at $C=1$ to 1270-1300~ms at $C=40$, while ML-DSA-based configurations remain within approximately 110-1600~ms across the same range. Classical configurations (P-256 and X25519) exhibit similar scaling, reaching approximately 1200-1500~ms at $C=40$.
Latency growth is approximately linear up to moderate concurrency ($C \leq 20$), after which deviations appear for computationally intensive schemes, particularly SLH-DSA, where latency increases more sharply.

\subsubsection{Energy Consumption Characteristics}
Energy consumption per connection remains relatively stable at low and moderate concurrency but diverges across cryptographic schemes under higher load. SLH-DSA-based configurations typically consume around 200-230~mJ per connection at moderate concurrency, increasing to over 320~mJ in the worst-case configuration (HQC + SLH-DSA at $C=40$). In contrast, Falcon and ML-DSA-based configurations maintain lower and more stable energy consumption, generally within 135-155~mJ per connection across all concurrency levels, closely matching classical configurations.

A reduction in per-connection energy is observed as concurrency increases from $C=1$ to $C=4$, reflecting improved multi-core utilization and a more even distribution of fixed system overhead across concurrent handshakes. Beyond this point, energy remains largely stable across most configurations, with increases at $C=40$ indicating contention and system saturation.

\begin{figure}[htbp]
    \centering
    \includegraphics[width=\linewidth]{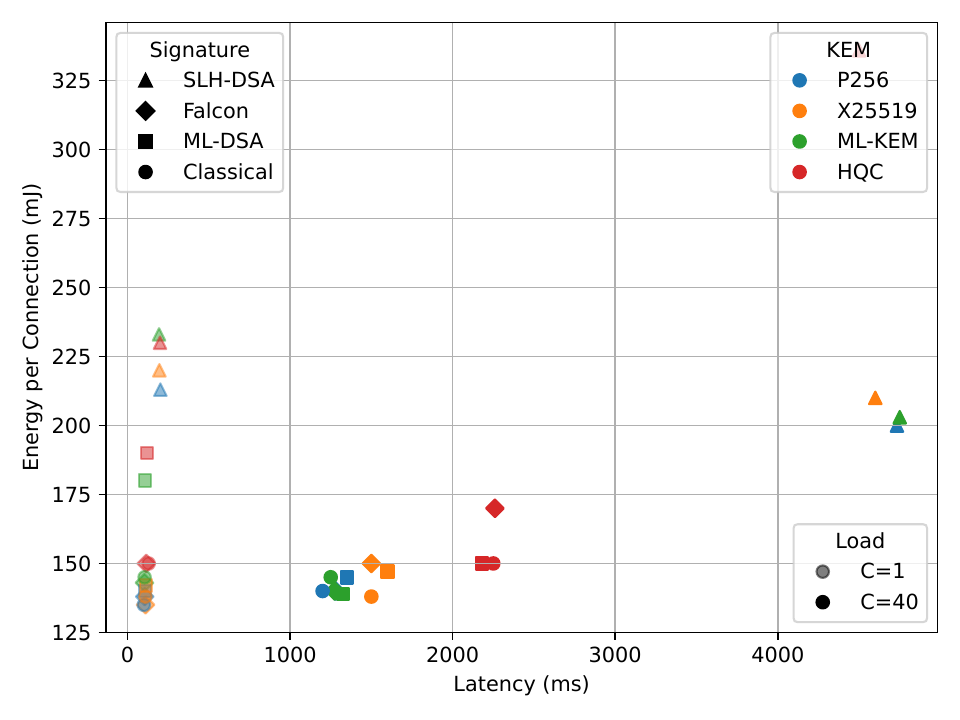}
    \caption{Latency-energy relationship for classical and post-quantum TLS configurations at low ($C=1$) and high ($C=40$) concurrency. Marker shape represents the signature schemes, and color denotes the KEMs.}
    \label{fig:latency_energy}
\end{figure}

Energy consumption is strongly correlated with latency, as higher-latency configurations incur greater total energy due to longer execution time. Fig.~\ref{fig:latency_energy} illustrates this relationship, where SLH-DSA-based configurations form a distinct high-latency, high-energy cluster, while Falcon and ML-DSA-based schemes remain closer to classical performance.

\subsubsection{Computational Load and Resource Utilization}

CPU utilization varies significantly across configurations. SLH-DSA-based configurations exhibit the highest CPU utilization, exceeding 30\% at high concurrency (e.g., approximately 36\% at $C=40$), while Falcon and ML-DSA-based configurations remain below 2-3\%. Memory usage was monitored throughout all experiments and remained stable and well within available capacity (e.g., within approximately 5.8-6.2~MB) across every configuration and concurrency level, confirming that observed performance differences are not memory-bound. Detailed memory traces are omitted from the tables for brevity. CPU values are averaged over the full experiment interval across all cores; short cryptographic bursts can therefore produce low average percentages even when they dominate per-handshake latency.

\subsubsection{Throughput Behavior and Saturation}

Throughput increases with concurrency up to a saturation point. For example, throughput increases from approximately 200-250~KB/s at $C=1$ to approximately 440-460~KB/s at $C=20$ for SLH-DSA-based configurations.
Beyond $C=20$, throughput plateaus, with minimal improvement at $C=40$. This indicates that the system reaches its maximum cryptographic processing capacity at moderate concurrency levels.

\subsubsection{Network Behavior and Stability}
Network-related metrics remain stable across all configurations. RTT varies from approximately 4.7~ms to 18.0~ms across all concurrency levels, with no dependence on the cryptographic scheme. The significant disparity between handshake latency and RTT indicates a shift in the dominant performance bottleneck. While RTT remains stable across all concurrency levels, latency increases with both the complexity of the cryptographic primitives and the number of concurrent client UEs. This suggests that the primary constraint in post-quantum TLS is computational processing rather than network transmission overhead, with system-level contention under concurrency further amplifying the effect.
Retransmissions remain negligible at low concurrency and increase under high load, indicating increasing processing delays rather than network instability.

\subsubsection{Concurrency-Induced Stress and System Degradation}
\begin{figure*}[htbp]
    \centering
    \includegraphics[width=0.7\linewidth]{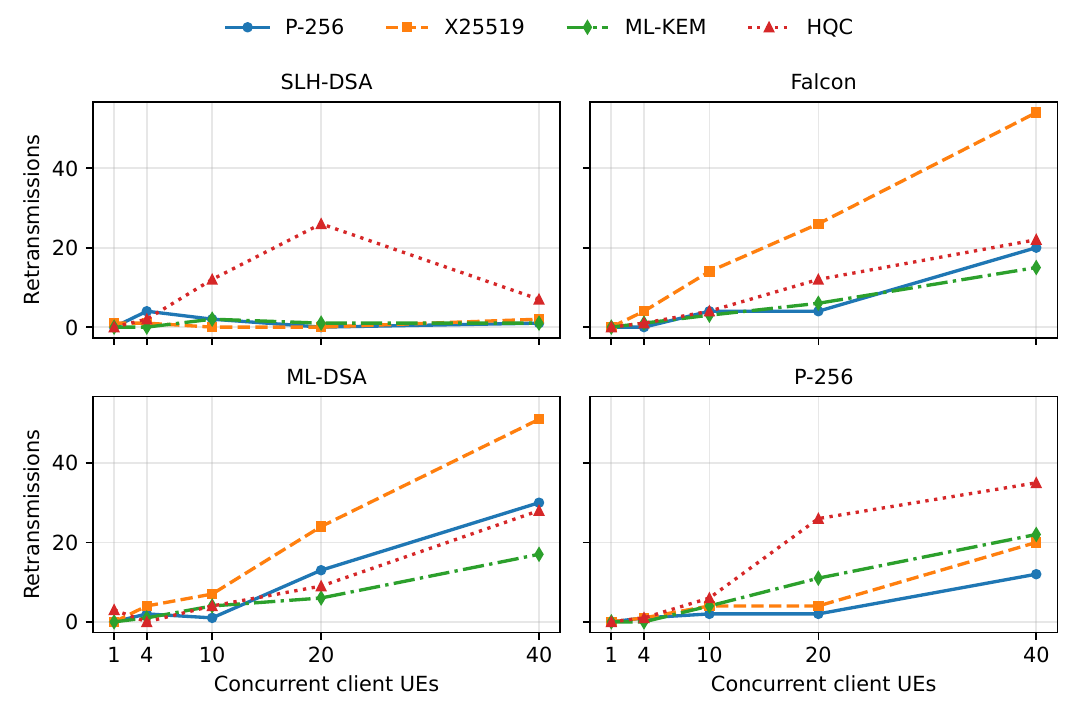}
    \caption{Retransmissions versus concurrent client UEs across signature schemes and KEMs. Each subplot shows a signature scheme, with lines denoting KEMs. Retransmissions increase with concurrency, indicating growing delays. The drop for HQC + SLH-DSA reflects saturation, where fewer retransmissions result from incomplete handshakes rather than improved performance.}
    \label{fig:retx}
    \vspace{-0.2in}
\end{figure*}

As concurrency increases, retransmissions rise across most configurations, as shown in Fig.~\ref{fig:retx}. For example, X25519 + Falcon increases from 0 retransmissions at $C=1$ to over 50 at $C=40$, while ML-DSA-based configurations exhibit similar monotonic growth.
SLH-DSA-based configurations show irregular behavior at high load. In particular, HQC + SLH-DSA increases to approximately 26 retransmissions at $C=20$, followed by a decrease to below 10 at $C=40$. This reduction corresponds to incomplete handshake execution (approximately 1315 out of 2000), indicating reduced processing capacity rather than improved performance. Fewer packets are transmitted, leading to fewer retransmissions.
This behavior reflects a transition from a congestion regime to a saturation regime, where it is unable to sustain the requested workload.


\subsubsection{Power Characteristics}

Server peak power remains within approximately 3.5-5~W across all configurations. Client peak power is significantly higher, reaching up to 7.5-8~W under high concurrency.
These peaks are transient and correspond to bursts of cryptographic processing and network activity.

\subsubsection{Thermal Behavior}

Server temperature increases from approximately 60-62$^\circ$C at $C=1$ to 68-70$^\circ$C at high concurrency, remaining below the 80$^\circ$C throttling threshold.
A thermal plateau is observed beyond $C \geq 20$, indicating steady-state operation. Client temperature remains approximately 5-7$^\circ$C lower than server temperature.

\subsubsection{Operating Regimes}

Three operating regimes are observed: \textit{(i) Low concurrency ($C \leq 4$):} low latency ($<120$~ms), minimal retransmissions, low energy consumption (e.g., $\sim$100 ms for ML-KEM + Falcon).
\textit{(ii) Moderate concurrency ($10 \leq C \leq 20$):} latency increases (300-700~ms for efficient schemes such as ML-DSA-based configurations), retransmissions rise, throughput approaches maximum.
\textit{(iii) High concurrency ($C=40$):} latency peaks (up to $\sim$4800~ms for SLH-DSA), throughput saturates, and incomplete handshake execution occurs.

\subsection{Design Implications}
\label{subsec:discussion}

\subsubsection{Computational Bottleneck and Network Decoupling}

The results reveal a clear decoupling between network performance and application-level performance. Despite stable RTT and minimal packet loss, latency increases significantly with concurrency, indicating that system performance is dominated by computational overhead rather than network limitations. 
From a design perspective, optimizing network infrastructure alone is insufficient to improve end-to-end performance for post-quantum TLS. Instead, system design must prioritize computational efficiency at the cryptographic layer.

\subsubsection{The Computational Wall of Hash-Based Signatures}

SLH-DSA configurations exhibit a sharp increase in latency under concurrency, reaching multi-second delays. This reflects the computational intensity of hash-based signature schemes, where repeated hash operations dominate execution time.
The observed saturation behavior indicates the presence of a computational limit beyond which additional concurrency does not improve throughput but instead increases delay. This suggests that hash-based signatures are not suitable for high-concurrency environments without hardware acceleration or workload offloading.

\subsubsection{Energy-Latency Coupling}

Energy consumption closely follows latency trends, demonstrating that execution time is the dominant factor in energy cost. Although instantaneous power remains bounded, prolonged execution results in higher total energy consumption.
This implies that reducing latency is directly aligned with improving energy efficiency. System designers should therefore prioritize algorithms and implementations that minimize execution time rather than focusing solely on reducing peak power.

\subsubsection{Power and Thermal Asymmetry}

A notable asymmetry exists between client and server behavior. The client experiences higher peak power due to burst-driven cryptographic operations, while the server exhibits sustained thermal load due to continuous processing.
The observed thermal plateau indicates that the system operates near a steady-state thermal limit. This suggests that sustained PQC workloads require careful thermal and power provisioning, particularly for continuously operating server nodes.

\subsubsection{Throughput Saturation and System Capacity}

Throughput saturation at moderate concurrency levels indicates a hard upper bound on system capacity. Beyond this point, additional concurrency increases latency and retransmissions without improving throughput.
This behavior implies that system scalability is constrained by cryptographic processing capacity. Effective system design must therefore consider concurrency limits and implement mechanisms such as load balancing or admission control to maintain performance.

\subsubsection{Implications for PQC Algorithm Selection}

Cryptographic design has a direct impact on system feasibility. Hash-based signatures introduce substantial computational and energy overhead, while lattice-based schemes achieve significantly better performance characteristics.
This indicates that algorithm selection is a critical design decision. For resource-constrained platforms, computationally efficient PQC schemes are essential to achieve acceptable performance and energy efficiency.

\subsubsection{Suitability for 5G Applications}

The evaluated configurations highlight limitations for latency-sensitive applications. Computationally intensive PQC schemes introduce substantial delays even under low concurrency, with latency increasing significantly under load.
This suggests that such schemes are not suitable for applications with strict latency requirements unless supported by hardware acceleration or architectural optimizations.

\subsubsection{System-Level Implications}
Overall, the results demonstrate that:
(i) system performance is CPU-bound rather than network-bound,  
(ii) energy consumption is primarily determined by execution time,  
(iii) throughput is limited by cryptographic processing capacity, and  
(iv) hardware constraints significantly influence PQC deployment.
These findings emphasize the need for co-design between cryptographic algorithms and system architecture to enable efficient and scalable deployment of post-quantum secure communication systems.

\section{Conclusion and Future Work}
\label{sec:conclusion}
This paper presented a system-level evaluation of classical and post-quantum TLS handshakes between embedded UE-class nodes over a disaggregated UERANSIM/Open5GS 5G user-plane path. The results show that performance is primarily limited by cryptographic computation and concurrency-induced contention rather than network transport effects. Signature choice is a key factor: hash-based signatures impose substantial computational and energy overhead, reducing scalability under load, while lattice-based schemes maintain more stable performance and energy behavior. As concurrency increases, the system progresses from efficient execution to congestion and saturation, where capacity is bounded by computational throughput. These findings highlight the need to select post-quantum primitives that align with the constraints of resource-limited 5G platforms.

Future work will explore hardware acceleration, broader device and deployment evaluations, protocol-level cost analysis, dynamic network conditions, over-the-air radio effects, and external shunt-based or dedicated power-analyzer measurements.



\section*{Acknowledgment}
This work was supported in part by the National Science Foundation under CAREER Award No.~2542642 and Grant No.~2552681.

\balance

\bibliographystyle{ieeetr}
\bibliography{References}
\end{document}